\documentclass[conference]{IEEEtran}

\usepackage{setspace}
\usepackage{caption}
\usepackage{subcaption}
\usepackage{longtable, array}
\usepackage{multirow}
\usepackage[titlenumbered,ruled]{algorithm2e}
\usepackage{amsmath}
\usepackage{amssymb}
\usepackage{amsthm}
\usepackage{lineno}
\modulolinenumbers[5]
\usepackage{url}

\newcolumntype{C}[1]{>{\centering\arraybackslash}p{#1}}
\newcolumntype{P}[1]{>{\centering\arraybackslash}p{#1}}
\newcolumntype{M}[1]{>{\centering\arraybackslash}m{#1}}

\usepackage{graphicx}
\DeclareGraphicsExtensions{.pdf,.jpeg,.png}

\begin{document}


\title{Affinity Network Fusion and Semi-supervised Learning for Cancer Patient Clustering}

\author{\IEEEauthorblockN{Tianle Ma}
	\IEEEauthorblockA{Department of Computer Science and Engineering\\
		University at Buffalo (SUNY)\\
		Buffalo, New York 14260-2500\\
		Email: tianlema@buffalo.edu}
	\and
	\IEEEauthorblockN{Aidong Zhang}
	\IEEEauthorblockA{Department of Computer Science and Engineering\\
		University at Buffalo (SUNY)\\
		Buffalo, New York 14260-2500\\
		Email: azhang@buffalo.edu}}





\maketitle
\begin{abstract}
Defining subtypes of complex diseases such as cancer and stratifying patient groups with the same disease but different subtypes for targeted treatments is important for personalized and precision medicine. 
Approaches that incorporate multi-omic data are more advantageous to those using only one data type for patient clustering and disease subtype discovery. However, it is challenging to integrate multi-omic data as they are heterogeneous and noisy. In this paper, we present Affinity Network Fusion (ANF) to integrate multi-omic data for patient clustering. ANF first constructs patient affinity networks for each omic data type, and then calculates a fused network for spectral clustering. We applied ANF to a processed harmonized cancer dataset downloaded from GDC data portal consisting of 2193 patients, and generated promising results on clustering patients into correct disease types. Moreover, we developed a semi-supervised model combining ANF and neural network for few-shot learning. In several cases, the model can achieve greater than 90\% acccuracy on test set with training less than 1\% of the data. This demonstrates the power of ANF in learning a good representation of patients, and shows the great potential of semi-supervised learning in cancer patient clustering. 
\end{abstract}




\section{Introduction}
Cancer patients are heterogeneous and complex. 
Patients with cancer from the same primary sites can be very different from each other in terms of disease progression, response to treatments, etc. One important task is to further cluster cancer patients of the same cancer type into subgroups and define new cancer subtypes with comprehensive molecular signatures associated with distinct clinical features. 

While the omic data collected are comprehensive, they are heterogeneous and noisy, too. If we use each type of omic data to cluster patients, we can probably generate different results. Since each type of omic data may contain some complementary information about the patients and the disease, we can perform clustering by integrating multi-omic data. Consensus clustering \cite{Monti2003} and its variants have been proposed to ``synthesize'' different clustering results. However, they only combine the final clustering results generated by different feature sets, and thus the integration is considered to be ``shallow''. Many methods that have been developed to integrate multi-omic data directly for patient clustering in the past several years are either based on probabilistic models or network models \cite{Bersanelli2016}. It has been demonstrated that patient clustering based on similarity network fusion (SNF) \cite{Wang2014} can often achieve promising results compared with other methods such as iCluster \cite{Shen2012} or KMeans. While SNF works well in clustering patients, we find that the required computational operations in SNF can be significantly reduced and simplified to get a reliable fused affinity network. 

Based on SNF, we developed Affinity Network Fusion (ANF) with several advantages. ANF requires much less computation while generating as good as or even better results than those from SNF. ANF provides a more general framework for complex object clustering with multi-view data, and can incorporate view weights. Moreover, we developed a semi-supervised model combining ANF and Neural Network (NN). Surprisingly, we can train less than 1\% of data and get greater than 90\% accuracy on test set using the output of ANF as input for a neural network classifier. 

We performed extensive experiments on a selected cohort of 2193 cancer patients from four primary sites and nine disease types. We have achieved high clustering/classification accuracy both by using ANF and spectral clustering without any training (i.e., zero-shot learning) and by training only a few labeled examples with our semi-supervised model (i.e., few-shot learning). The results demonstrate the power of ANF in learning good patient representations for clustering/classification purpose.

\subsection{Related Work} \label{sec:related_work}

Mining multi-omic data has been a hot topic in recent years \cite{Bersanelli2016,Meng2016}. 
Many approaches are based on probabilistic modeling usually involving a set of latent variables with a prior distribution. For instance, iCluster \cite{Shen2012}, a widely used cancer patient clustering method, assumes different types of omic data share a common latent feature space that can be jointly learned from multi-omic data. KMeans clustering is then performed on the learned latent features. 
Other approaches incorporate domain knowledge such as biological networks into probabilistic modeling. For instance, PARADIGM \cite{Vaske2010} converted NCI pathway databases into a factor graph in which each gene is a factor incorporating several kinds of information, a ``natural'' way for integrating multi-omic data. 

As it is challenging to define a proper prior distribution and learn a good posterior distribution with limited data, many other approaches do not use probabilistic modeling. For example, Similarity Network Fusion (SNF) \cite{Wang2014} constructs patient similarity networks using different types of omic features, and fuses multiple patient similarity networks to achieve a ``consensus'' network that is then used for clustering patients into disease subtypes.

Based on the main idea of SNF \cite{Wang2014}, we developed a simpler and more general framework, Affinity Network Fusion (ANF), to combine multiple networks into a fused consensus network. The fused network captures complementary information from multiple views and is much more robust to noise than individual networks learned from each view. 

\section{Affinity Network Fusion (ANF)} \label{sec:anf}

ANF differs from clustering methods that directly operate on patient-feature matrix, e.g., KMeans, in that it applies graph clustering to a constructed patient affinity/similarity matrix instead of patient-feature matrix.
Thus the success of ANF relies on the construction of an ``accurate'' patient affinity network that incorporates information from multiple views.

As multi-omic data are heterogeneous, we first construct a patient affinity network from each view (i.e., an -omic data source), and then fuse all individual networks to get a more robust one. In order to make patient affinity network robust to noise, we mainly employ two nonlinear k-Nearest-Neighbor(kNN)-based transformations: kNN Gaussian kernel and kNN graph. In the following we briefly describe the entire network construction and fusion process from raw patient-feature matrices to fused patient affinity network.

%

\subsection{Construct Patient kNN Affinity Networks}

Suppose there are $N$ patients and $n$ views (i.e., omic feature spaces). 
Let $\mathcal{X}^{(v)}, v = 1,2,\cdots, n$, be $n$ patient-feature matrices. For example, $\mathcal{X}^{(1)}$ might be a gene expression matrix with rows corresponding to patients (or samples) and columns corresponding to gene expression measurements (e.g., FPKM values). 
Since omic data are usually high dimensional and noisy, feature selection and transformation \cite{Ma2017,Ma2016} is often needed to transform raw patient-feature matrices $\mathcal{X}^{(v)}, v=1,2,\cdots, n$ into processed patient-feature matrices $\mathbf{X}^{(v)}, v=1,2,\cdots, n$. 

One simple way to integrate multi-view data is simply concatenating all individual feature matrices $\mathbf{X}^{(v)}, v=1,2,\cdots, n$ into a single matrix. However, features from different views may follow different distributions and have different scales. It is possible that features from some views dominate others, making the integration less effective. In our approach, we first process individual views separately to generate patient affinity networks and then combine them through network fusion, thus avoiding dealing with heterogeneous feature spaces directly.

For each view $v$, we first calculate a pair-wise patient distance matrix $\mathbf{\Delta}^{(v)}=(\delta_{ij}^{(v)})_{N\times N}$ based on processed patient-feature matrix $\mathbf{X}^{(v)}$ using a certain distance metric such as Euclidean distance, and then calculate kNN Gaussian kernel and kNN graph to get patient affinity matrix.

\subsubsection{kNN Gaussian Kernel} \label{sec:knn}


With pair-wise distance matrix $\mathbf{\Delta}^{(v)}$, we can construct corresponding patient similarity network $\mathbf{S}^{(v)}$ in multiple ways. For example \cite{Luxburg2007}, 

\begin{itemize}
	\item $\epsilon$-neighborhood graph (unweighted): connect two patients if and only if the distance between them is smaller than $\epsilon$. The choice of $\epsilon$ is problem-dependent.
	
	\item $k$-nearest-neighbor graph (unweighted): Connect each patient with his or her $k$ nearest neighbors. $k$ is a parameter to tune. 
	
	\item Fully connected graph (weighted): connect all patients with weighted edges.  Edge weights can be calculated using some kernel such as Gaussian kernel. 
\end{itemize}

Here we combine local Gaussian kernel and kNN graph to define a kNN Gaussian kernel (the idea is from \cite{Wang2014}):

\begin{equation}
\label{eq:mu_i}
\mu_i=\frac{\sum_{l\in \mathcal{N}_k(i)}{\delta_{il}}}{k}
\end{equation}

\begin{equation}
\label{eq:sigma_ij}
\sigma_{ij}=\alpha(\mu_i+\mu_j) + \beta \delta_{ij}
\end{equation}

\begin{equation}
\label{eq:k_ij}
K_{ij}=\frac{1}{\sqrt{2\pi}\sigma_{ij}}e^{-\frac{\delta_{ij}^2}{2\sigma_{ij}^2}}
\end{equation}

In Eq.~\ref{eq:mu_i}, Eq.~\ref{eq:sigma_ij} and Eq.~\ref{eq:k_ij}, we omit the superscript (i.e., $\cdot ^{(v)}$) for simplicity (For example, $\delta_{ij}$ represents $\delta_{ij}^{(v)}$, the distance between patient $i$ and $j$ calculated from view $\mathbf{X}^{(v)}$). 
In Eq.~\ref{eq:mu_i}, $\mathcal{N}_k(i)$ represents the indexes of k-nearest neighbors of patient $i$. The choice of $k$ is important and needs to be tuned. $\mu_i$  represents the local diameter of node $i$. $\sigma_{ij}$ in Eq.~\ref{eq:sigma_ij} incorporates both local diameters of patient $i$ and $j$ and their distance. Eq.~\ref{eq:k_ij} calculates kNN Gaussian kernel between patient $i$ and $j$, with $\sigma_{ij}$ defined as Eq.~\ref{eq:sigma_ij}, to incorporate local network structure. 

\subsubsection{kNN Graph}

We can regard $K_{ij}^{(v)}$ (Eq.~\ref{eq:k_ij}) as an unnormalized similarity measure between patient $i$ and $j$ from view $v$. We further normalize $K_{ij}^{(v)}$ to $S_{ij}^{(v)}$ by Eq.~\ref{eq:s_ij}. As each row of $\mathbf{S}^{(v)}=(S_{ij}^{(v)})_{N\times N}$ sums to 1, $S_{ij}^{(v)}$ can be regarded as normalized similarity measure between patient $i$ and $j$ in view $v$, or the probability of (the state of) patient $i$ transitions to (the state of) patient $j$ in view $v$. 

\begin{equation}
\label{eq:s_ij}
S_{ij}^{(v)} = \frac{K_{ij}^{(v)}}{\sum_{j=1}^{N}K_{ij}^{(v)}}, \quad 1 \le i,j \le N
\end{equation}

\paragraph{\textbf{Further Prune ``Weak'' Edges}}
$\mathbf{S}^{(v)}$ represent a fully connected affinity network (with positive edges between every patient pairs). Since edges with small weights are more likely to be noise, we prune ``weak'' edges by constructing a kNN graph from $\mathbf{S}^{(v)}$ (Eq.~\ref{eq:knn_transition}). 

\begin{equation}
\label{eq:knn_transition}
W_{ij}^{(v)}=
\begin{cases}
(1-\epsilon)\frac{S_{ij}^{(v)}}{\sum_{j\in \mathcal{N}_k(i)}S_{ij}^{(v)}}, & \text{if}\ j \in \mathcal{N}_k(i) \\
\epsilon \frac{S_{ij}^{(v)}}{\sum_{j\notin \mathcal{N}_k(i)}S_{ij}^{(v)}}, & \text{otherwise}
\end{cases}
\end{equation}

In Eq.~\ref{eq:knn_transition}, $\mathcal{N}_k(i)$ again refers to the indexes of $k$ nearest neighbors of patient $i$. $\epsilon$ is a very small number. We can set $\epsilon = 0$, then for each row of $\mathbf{W}^{(v)}$, only $k$ elements are non-zero, and only the weights of $k$ nearest neighbors are used for normalization. 

In fact $\mathbf{W}^{(v)}$ can be seen as a trunked version of $\mathbf{S}^{(v)}$ by ``reducing or throwing away'' weak signals (i.e., small edge weights) in $\mathbf{S}^{(v)}$. Thus $\mathbf{W}^{(v)}$ should be more robust to small noise. Since each row of $\mathbf{W}^{(v)}$ sums to 1, it can also be regarded as a state transition matrix.

\subsection{Fuse Multiple Affinity Networks}



Suppose there are $n$ views to be combined. 
Let $\mathbf{w}=(w_1, w_2,\cdots, w_n)$ be the view weights. (If the view weights are not given, we usually start with uniform weights, and tune the weights with semi-supervised learning.) We can calculate a weighted view by Eq.~\ref{eq:fused_mat}

\begin{equation}
\label{eq:fused_mat}
\begin{split}
&\mathbf{W} = \sum_{v=1}^{n}w_v\cdot \mathbf{W}^{(v)}\\
&\sum_{v=1}^{n}w_v=1, \quad w_v \ge 0
\end{split}
\end{equation}

As $\mathbf{W}^{(v)}$ is row-normalized (Eq.~\ref{eq:knn_transition}), the fused view $\mathbf{W}$ (Eq.~\ref{eq:fused_mat}) is also row-normalized (each row sums to 1), and can be regarded as a state transition matrix. In Eq.~\ref{eq:random_walk}, we multiply $\mathbf{W}$ by itself $r$ times, and get a smoothed version of transition matrix. This process can be interpreted as $r$-step random walk on a patient affinity network.

\begin{equation}
	\label{eq:random_walk}
	\mathbf{W}^{*} = \mathbf{W}^{r}
\end{equation}

In Eq.~\ref{eq:random_walk}, if $r$ is large enough, $\mathbf{W}^{*}$ will reach certain stationary point \cite{Luxburg2007} and become rank 1 (all rows become the same vector). However, it is not desirable for $r$ to be too large. We can achieve good results in our experiments on TCGA data by setting $r=1$ (simply Eq.~\ref{eq:fused_mat}) and $r=2$. If we keep increasing $r$, the results will not change dramatically in the beginning, but will deteriorate when $r$ becomes too large (e.g., $r>10$).

In the following we will discuss a slightly modified formulation of Eq.~\ref{eq:random_walk} when $r=1,2$. 

\subsubsection{Alternative Formulation of Affinity Network Fusion for Special Cases}

In Eq.~\ref{eq:fused_mat}, we only used $\mathbf{W}^{(v)}$ to get a aggregated view. Sometimes the pruned weak edges might be useful, too. In Eq.~\ref{eq:anf1}, we include $\mathbf{S}^{(v)}$ as well. 

\begin{equation}
\label{eq:anf1}
\begin{split}
\mathbf{W}^{(v)} = \beta_1 \mathbf{W}^{(v)} &+ \beta_2 \overline{\mathbf{W}^{(-v)}} + \beta_3 \mathbf{S}^{(v)} + \beta_4 \overline{\mathbf{S}^{(-v)}}\\
&\sum_{v=1}^{4}\beta_v = 1, \beta_v \ge 0
\end{split}
\end{equation}

\begin{equation}
\label{eq:complementary_view}
\begin{split}
\overline{\mathbf{W}^{(-v)}} &= \sum_{i\ne v} \frac{w_i}{\sum_{j\ne v}w_j} \cdot \mathbf{W}^{(i)}\\
\overline{\mathbf{S}^{(-v)}} &= \sum_{i\ne v}\frac{w_i}{\sum_{j\ne v}w_j} \cdot \mathbf{S}^{(i)}
\end{split}
\end{equation}

In Eq.~\ref{eq:anf1}, $\overline{\mathbf{W}^{(-v)}}$ and $\overline{\mathbf{S}^{(-v)}}$ represent a weighted complementary view from $n-1$ other views as defined in Eq.~\ref{eq:complementary_view}. Eq.~\ref{eq:anf1} can be interpreted as network diffusion between view $v$ and other complementary views, resulting in a smoother version of $\mathbf{W}^{(v)}$.
In practice, since $\mathbf{W}^{(v)}$ is usually more robust to noise than $\mathbf{S}^{(v)}$, we often set $\beta_3 = \beta_4 = 0$, in which case Eq.~\ref{eq:anf1} is equivalent to Eq.~\ref{eq:fused_mat}.

Inspired by Similarity Network Fusion \cite{Wang2014}, we can define a more complex fusion process as Eq.~\ref{eq:anf2}.

\begin{align}
\label{eq:anf2}
\begin{split}
\mathbf{W}^{(v)} = &\alpha_1 \mathbf{W}^{(v)}\cdot \overline{\mathbf{W}^{(-v)}} +
\alpha_2 \overline{\mathbf{W}^{(-v)}} \cdot \mathbf{W}^{(v)} +\\  
&\alpha_3 \mathbf{W}^{(v)}\cdot \overline{\mathbf{S}^{(-v)}} +
\alpha_4 \overline{\mathbf{S}^{(-v)}} \cdot \mathbf{W}^{(v)} + \\
&\alpha_5 \mathbf{S}^{(v)}\cdot \overline{\mathbf{W}^{(-v)}} +
\alpha_6 \overline{\mathbf{W}^{(-v)}} \cdot \mathbf{S}^{(v)} +\\ 
&\alpha_7 \mathbf{S}^{(v)}\cdot \overline{\mathbf{S}^{(-v)}} +
\alpha_8 \overline{\mathbf{S}^{(-v)}} \cdot \mathbf{S}^{(v)}
\end{split}
\end{align}
\begin{center}
	$\sum_{i=1}^{8}\alpha_i = 1, \alpha_i \ge 0$
\end{center}

In Eq.~\ref{eq:anf2}, $\overline{\mathbf{W}^{(-v)}}$ and $\overline{\mathbf{S}^{(-v)}}$ are defined the same as Eq.~\ref{eq:complementary_view}. 
Eq.~\ref{eq:anf2} is roughly equivalent to Eq.~\ref{eq:random_walk} when $r=2$. The first term $\alpha_1 \mathbf{W}^{(v)}\cdot \overline{\mathbf{W}^{(-v)}}$ can be interpreted as a two-step random walk (by multiplying two transition matrices): the first step is a random walk on view $v$, and the second step is a random walk on the aggregated complementary view ($\overline{\mathbf{W}^{(-v)}}$, Eq.~\ref{eq:complementary_view}). The other terms in Eq.~\ref{eq:anf2} can have similar interpretations.

Our experiments on cancer genomic data show that the terms using $\mathbf{W}^{(v)}$ usually works better than using $\mathbf{S}^{(v)}$, suggesting $\mathbf{W}^{(v)}$ is more reliable than $\mathbf{S}^{(v)}$. In practice, the default choice is just using the first two terms:

\begin{equation}
\label{eq:anf2-simple}
\mathbf{W}^{(v)} = \alpha \mathbf{W}^{(v)}\cdot \overline{\mathbf{W}^{(-v)}} +
(1-\alpha) \overline{\mathbf{W}^{(-v)}} \cdot \mathbf{W}^{(v)}
\end{equation}

Note both Eq.~\ref{eq:anf1} and Eq.~\ref{eq:anf2} can be seen as recursive formulas. We can iteratively update $\mathbf{W}^{(v)}$ until convergence (usually within a few iterations). However, in practice, it is not necessary to require convergence, which can be seen as a stationary state achieved through a sufficient number of random walks. In fact, we only need to calculate Eq.~\ref{eq:anf1} (which can be seen as a one-step random walk) and Eq.~\ref{eq:anf2} (which can be seen as a two-step random walk) once to get ``smoothed'' views, and calculate a weighted average of all views as shown in Eq.~\ref{eq:fused_mat}. This is one major difference from SNF\cite{Wang2014}, which requires a number of iterations to update similarity matrix through multiple matrix multiplications until convergence. By contrast, ANF essentially only needs one iteration to get as good as or even better results than SNF \cite{Ma2017}.

\paragraph{Interpretation} Each disease subtype may have its own molecular signature, which can be represented by a state space model with a vector of explicit and latent variables. Disease subtype discovery is essentially finding a unique state representation for each subtype. Ideally, patients belonging to a disease subtype should fall into its corresponding space, and be near each other, while patients with different disease subtypes should be far away in their state space representations.

All $\mathbf{W}^{(v)}$ and $\mathbf{S}^{(v)}$ can be regarded as state transition matrices learned from multi-view data, and Eq.~\ref{eq:fused_mat} essentially computes a weighted state transition matrix $\mathbf{W}$. As patients with the same disease subtypes will be more likely to stay in the same state space, Eq.~\ref{eq:random_walk} and Eq.~\ref{eq:anf2} will strength the network modularity: patients belonging to the same group will have denser edges, while the edges among patients from different groups will be sparser. The resulted fused network (Eq.~\ref{eq:fused_mat}) will thus be well suited for graph clustering with its modular structures corresponding to disease subtypes.

\paragraph{Toy example} 

\begin{figure}[!t]
	\centering
	\includegraphics[width=0.5\textwidth]{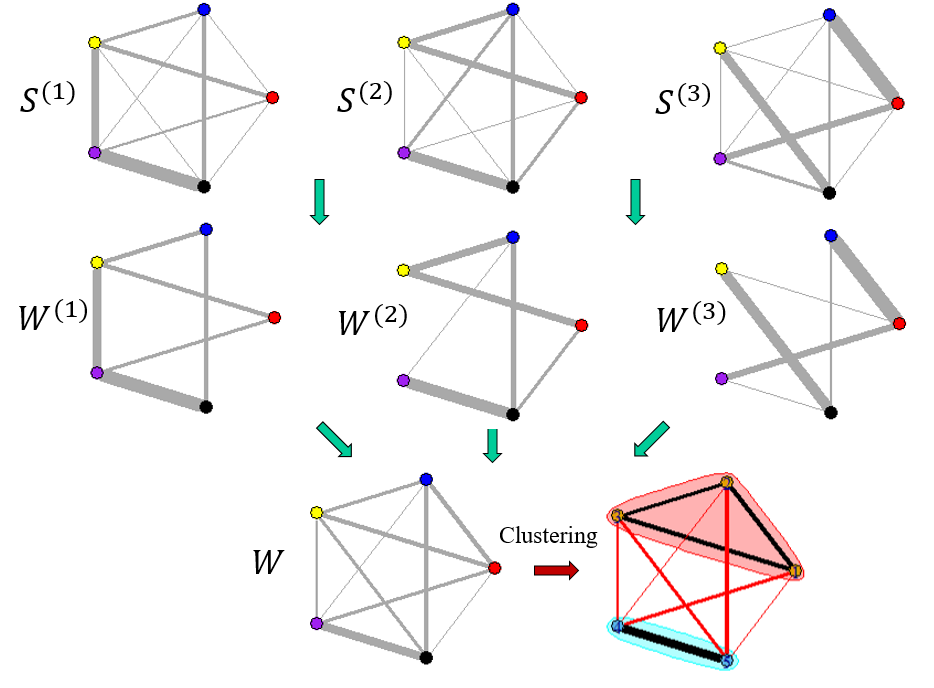}
	\caption{Toy example of ANF}
	\label{fig:toy_anf}
\end{figure}

In order to give a vivid understanding of the clustering process, we show a toy example in Fig.~\ref{fig:toy_anf}.
In this figure, the nodes with different colors represent five patients.
The edges represent transition probabilities between (the states of) two patients. The width of an edge shows how strong the edge is (i.e., edge weight).
The first row corresponds to state transition matrix $\mathbf{S}^{(v)}$ calculated from Eq.~\ref{eq:s_ij}. 
The second row corresponds to kNN transition matrix $\mathbf{W}^{(v)}$ (trunked version of $\mathbf{S}^{(v)}$). 
We can see that the ``noisy'' edges in $\mathbf{S}^{(v)}$ have been removed from $\mathbf{W}^{(v)}$.
The third row shows the fused $\mathbf{W}$, which combines information from all three views. 
The goal of spectral clustering is to minimize the total effect of between-cluster edges, which can be mathematically defined as normalized graph cut \cite{Luxburg2007}. The spectral clustering result on this five nodes toy example is shown in the bottom right subfigure. As we can see three nodes are clustered together (red shaded area), and the other two are clustered (lightblue shaded area). 
The edges within each cluster are relatively stronger compared with the edges across clusters (red edges in the figure). 

The overall unsupervised ANF framework for cancer patient clustering (zero-shot learning) is summarized in Alg.~\ref{alg:ANF_framework}.

\begin{algorithm}
	
	\SetKwInOut{Input}{Input}
	\SetKwInOut{Output}{Output}
	
	\Input{\textbullet{Patient-feature matrices ($n$ views): $\mathcal{X}^{(v)},v=1,2,\cdots,n$} \\ 
		\textbullet{Number of clusters: $c$}\\
		\textbullet{Weight of each view (optional): $\mathbf{w}=(w_1, \cdots, w_n)$} \\
		\textbullet{Other optional parameters}}
	\Output{
		\textbullet{Patient cluster assignment $\mathcal{A}$}\\
		\textbullet{Fused patient affinity matrix $\mathbf{W}$}\\
		\textbullet{Patient affinity matrices from each view, $\mathbf{W}^{(v)}, v=1,\cdots, n$}
		}
	\Begin{\textbf{Feature selection and transformation}\\
		$\mathcal{X}^{(v)} \rightarrow \mathbf{X}^{(v)}\in \mathbb{R}^{N\times p_v},v=1,2,\cdots,n$\\ 
		\textbf{Calculate pair-wise distance matrix for each view: $\mathbf{\Delta}^{(v)}\in \mathbb{R}_+^{N\times N},v=1,2,\cdots,n$}\\
		\textbf{Calculate kNN affinity matrix for each view: $\mathbf{W}^{(v)},v=1,2,\cdots,n$} (Eq.~\ref{eq:anf1} or Eq.~\ref{eq:anf2})\\
		\textbf{Calculate fused affinity matrix $\mathbf{W}$} (Eq.~\ref{eq:fused_mat} or Eq.~\ref{eq:random_walk})\\	
		\textbf{Spectral clustering on fused affinity matrix $\mathbf{W}$: $(\mathbf{W},c)\rightarrow \mathcal{A}$} \\
		\textbf{Return $\mathcal{A},\mathbf{W}, \mathbf{W}^{(v)},v=1,2,\cdots,n$}
	}	
	\caption{Unsupervised Affinity Network Fusion for Clustering}
	\label{alg:ANF_framework}
\end{algorithm}

\section{Semi-supervised Learning on Patient Affinity Networks}
In addition to being a principled method for clustering, ANF framework (Alg.~\ref{alg:ANF_framework}) generates a series of patient kNN affinity matrices $\mathbf{W}^{(v)}, v=1,2,\cdots, n$, which turn out to be good representations of patients for the purpose of classification or clustering. 
We have designed the following semi-supervised neural network model to tap the representation power of patient affinity matrices. 

\begin{figure}[!t]
	\centering
	\includegraphics[height=3.5in]{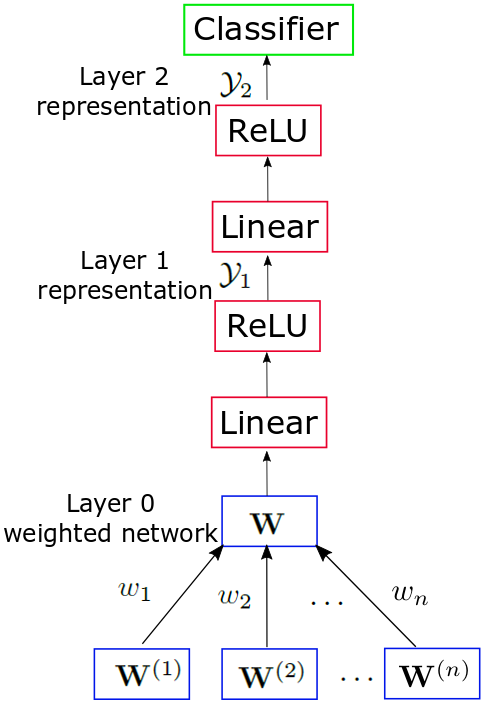}
	\caption{Semi-supervised clustering with neural network model}
	\label{fig:nn-classifier}
\end{figure}


\begin{align}
\label{eq:nn_classifier}
\begin{split}
\mathbf{W} &= \sum_{v=1}^{n}w_v\cdot \mathbf{W}^{(v)} \\
\mathcal{Y}_1 &= f(\mathbf{A_1 W+b_1}) \\
\mathcal{Y}_2 &= f(\mathbf{A_2 \mathcal{Y}_1+b_1})
\end{split}
\end{align}

where $f(x)=ReLU(x)=max(x, 0)$ or other nonlinear activations.

As shown in Fig.~\ref{fig:nn-classifier} and Eq.~\ref{eq:nn_classifier}, the input layer is the concatenated kNN affinity matrices for each view generated by Alg.~\ref{alg:ANF_framework}. \textbf{Layer 0} calculates a weighted fused view (similar to attention mechanism). In Eq.~\ref{eq:fused_mat}, we have to manually set view weights (usually uniformly). Now we can train a neural network end-to-end to learn these weights automatically.

\textbf{Layer 1} extracts the hidden representation $\mathcal{Y}_1$ from $\mathbf{W}$, and \textbf{Layer 2} produces a much lower dimensional hidden representation $\mathcal{Y}_2$ for classification.   

The reason we do not want more layers or add advanced modules such as skip connections \cite{He2016} is that ``class label'' information is very expensive in biomedical applications and we usually have very few training examples. Adding more layers will increase the risk of overfitting for a few training examples. Importantly, implicit kNN-based nonlinear transformations in Alg~\ref{alg:ANF_framework} have already been applied to the original feature matrices. The input to the neural network, kNN affinity matrices generated by Alg~\ref{alg:ANF_framework}, are already a good representation of the data. Though very simple, this architecture works surprisingly well (see Sec.~\ref{sec:res}).

\paragraph{How to Find Examples}
While it is costly to assign class label to all patients, it is feasible to label some patients as belonging to different groups based on clinical data. For example, as most cancer patient clustering methods used clinical data to validate the clustering results, we can also use clinical data to ``manually'' choose some examples. We call examples chosen using clinical data by experts as ``clean'' examples. On the other hand, we can select a few examples with cluster labels generated by unsupervised ANF framework (Alg.~\ref{alg:ANF_framework}) for training. We call examples with clustering labels generated by unsupervised learning methods as ``noisy examples'' since they may not represent true class labels.

\section{Experimental Results} \label{sec:res}

\subsection{Dataset and Evaluation Metrics}
Harmonized cancer datasets were downloaded from Genomic Data Commons Data Portal (\url{https://portal.gdc.cancer.gov/}).
We selected patients with cancers from four cancer primary sites: adrenal gland, lung, kidney, and uterus. Cancers from each of these primary sites have more than one disease types. For example, cancers from adrenal gland has two disease types: Pheochromocytoma and Paraganglioma (project name: TCGA-PCPG) and Adrenocortical Carcinoma (project name: TCGA-ACC). In this paper, for ease of description, we refer to ``cancer types'' as cancers from these four primary sites. We want to cluster patients of the same ``cancer types'' into known disease types. Since the disease types are one-to-one corresponded to TCGA projects. For ease of description, in the following we use TCGA project names to refer to disease types.
 
The number of samples used for analysis in each cancer type is summarized in Table~\ref{tbl:sample_info_4types} (a few ``outlier'' samples detected by exploratory data analysis had already been removed). All these patient samples have gene expression, miRNA expression and DNA methylation (from HumanMethylation450 array) data available for both tumor and normal samples.   

\begin{table}[t]
	\begin{center}
		\caption{Sample information of four cancer types}
		\label{tbl:sample_info_4types}
		\begin{tabular}{c|cc|c}
			\firsthline
			Cancer type &  \multicolumn{2}{c|}{Disease type} & Total \\
			\hline
			\multirow{2}{*}{adrenal gland} & TCGA-ACC & 76 & \multirow{2}{*}{253}\\
			& TCGA-PCPG & 177 &\\
			\hline
			\multirow{2}{*}{lung} & TCGA-LUAD & 447 & \multirow{2}{*}{811}\\
			& TCGA-LUSC & 364 &\\
			\hline
			\multirow{3}{*}{kidney} & TCGA-KICH & 65 & \multirow{3}{*}{654}\\
			& TCGA-KIRC & 316 & \\
			& TCGA-KIRP & 273 & \\
			\hline
			\multirow{2}{*}{uterus} & TCGA-UCEC & 421 & \multirow{2}{*}{475}\\
			\cline{2-3}
			& TCGA-UCS  & 54 & \\
			\lasthline
		\end{tabular}
	\end{center}
\end{table}

While our ultimate goal is to detect cancer subtypes (the true subtypes are not known), it is a good strategy to evaluate disease subtype discovery methods using a dataset with groundtruth. The dataset used here serves for this purpose well.  
Since we have ground truth disease types, we can evaluate clustering results using external metrics. 
The metrics we used to evaluate clustering results include: 
(1) Normalized Mutual Information (NMI): 
$NMI(\Omega, \mathcal{C})=\frac{I(\Omega, \mathcal{C})}{(H(\Omega)+H(\mathcal{C}))/2}$;
(2) Adjusted Rand Index (ARI) \cite{Hubert1985}.

\subsection{The Power of Affinity Network Fusion (ANF)}
In this section we mainly examine the performance of unsupervised ANF (Alg.~\ref{alg:ANF_framework}) without using any labeled data (i.e., zero-shot learning).

\begin{figure}[!t]
	\centering
	\includegraphics[width=4in]{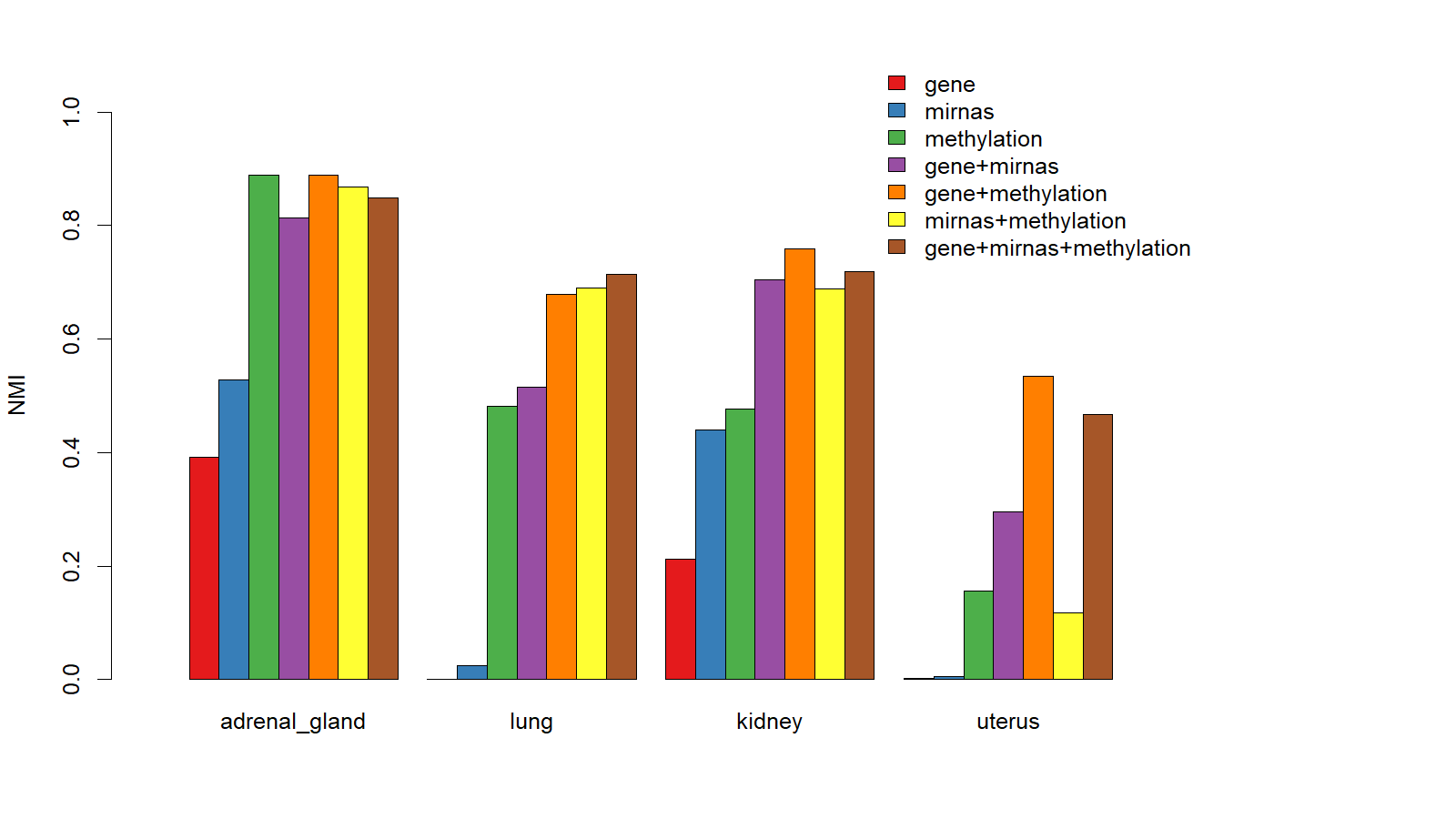}
	\caption{Power of ANF combining multi-omic data}
	\label{fig:power_ANF}
\end{figure}

To demonstrate the power of ANF, we compared the clustering results using single data types with those using ANF to integrate multiple data types.
In Fig.~\ref{fig:power_ANF}, we compared seven different combinations of data types: 
\begin{itemize}
	\item ``gene'': gene expression (FPKM values)
	\item ``mirnas'': miRNA expression (normalized counts)
	\item ``methylation'': DNA methylation data (beta values from Illumina Human Methylation 450 platform)
	\item ``gene+mirnas'': combine ``gene'' and ``mirnas'' using ANF
	\item ``gene+methylation'': combine ``gene'' and ``methylation'' using ANF
	\item ``mirnas+methylation'': combine ``mirnas'' and ``methylation'' using ANF
	\item ``gene+mirnas+methylation'': combine ``gene'', ``mirnas'', and ``methylation'' using ANF
\end{itemize}

Fig.~\ref{fig:power_ANF} shows NMI values (between 0 and 1. Larger NMI value corresponds to better clustering result) of patient clustering results using ANF framework on the aforementioned seven combinations of data types. (Here we set the number of clusters to be the number of disease types.) 

In general, a combination of at least two data types usually yields better clustering results. 
Specifically, for uterus cancer, clustering using gene or miRNA expression data alone did a ``terrible'' job (NMI $\approx 0$). However, by integrating the two data types, the result improves significantly (NMI=0.30), which demonstrates the power of ANF. 

We also find that it is usually not the case that integrating three data types would generate better results than that from integrating two data types. One possible reason is that clustering using DNA methylation beta values performs much better than using FPKM and normalized miRNA expression values for all four cancer types, suggesting that DNA methylation data may contain highly relevant information about disease types. However, integrating more data types tends to make the results more robust as ``gene+mirna+methylation'' consistently performs relatively well across four cancer types, while two data type combinations may fail in some cases (for example, ``mirnas+methylation'' for uterus cancer does not yield good clustering results).
Very similar results are obtained for using Adjust Rand Index (ARI) as clustering metric (not shown here). 

ANF can achieve high clustering accuracies for all the four cancer types (Table~\ref{tbl:cluster_accuracy_4types}). Code and more comprehensive results can be found in \url{https://github.com/BeautyOfWeb/Clustering-TCGAFiveCancerTypes}. ANF has been accepted as a Bioconductor package (\url{https://bioconductor.org/packages/release/bioc/html/ANF.html}).

\begin{table}[t]
	\begin{center}
		\caption{Clustering accuracy of four cancer types}
		\label{tbl:cluster_accuracy_4types}
		\begin{tabular}{c|c|c|c|c}
			\firsthline
			& Adrenal gland & Lung & Kidney & Uterus \\
			\hline 
			NMI & 0.96 & 0.75 & 0.84 & 0.61\\
			\hline
			ARI & 0.98 & 0.83 & 0.91 & 0.78 \\
			
			\lasthline
		\end{tabular}
	\end{center}
\end{table}
 
\subsubsection{Determine the Number of Clusters} \label{sec:eigengap}

Since ANF framework applies spectral clustering to a fused affinity matrix, we can use eigengap heuristic or more advanced technique \cite{Wang2017} to determine the number of clusters and indirectly assess cluster quality. For simplicity, we only discuss eigengap heuristic here. 

\begin{figure}[!t]
	\centering
	\includegraphics[width=0.5\textwidth]{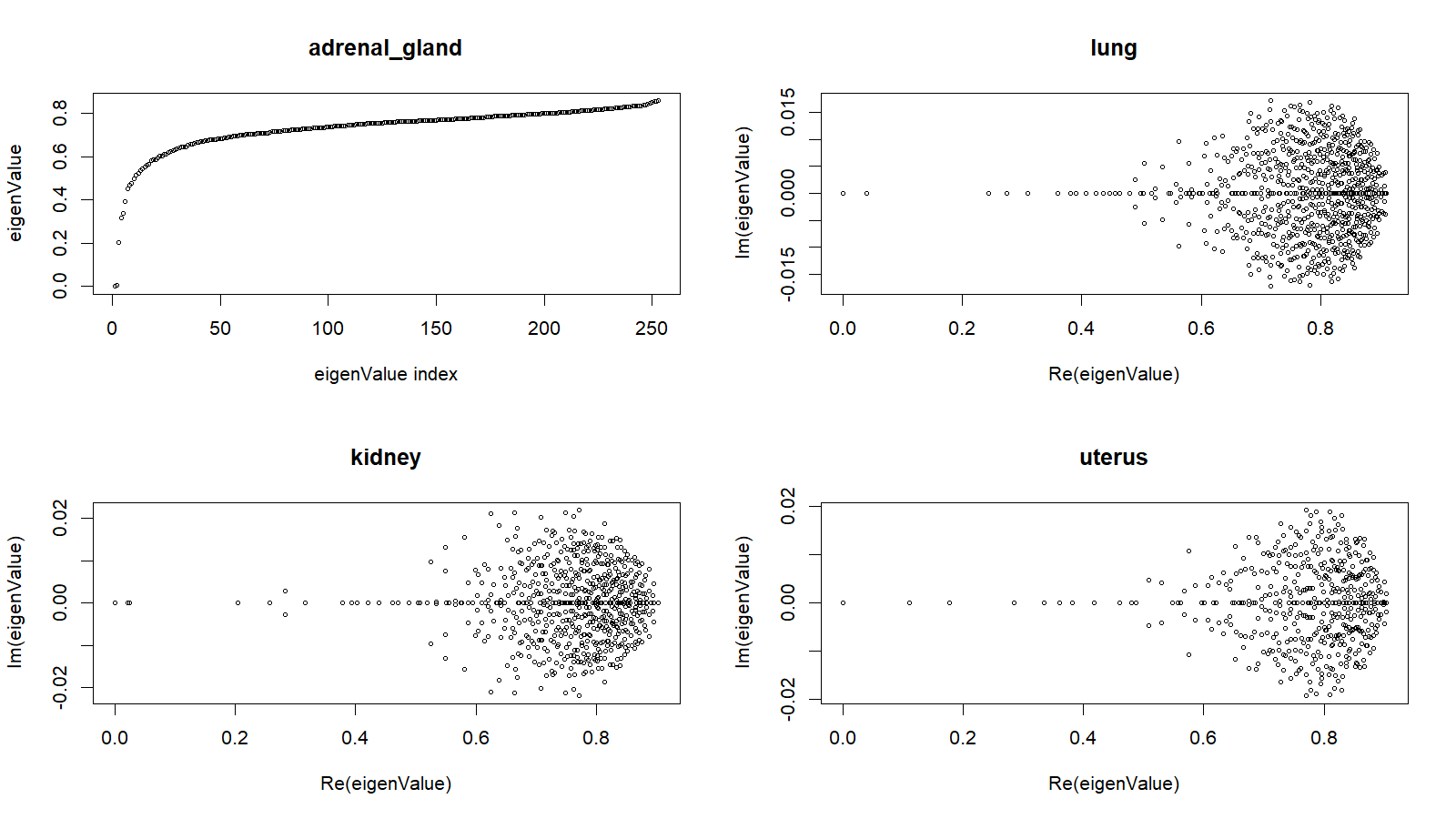}
	\caption{Eigenvalues of affinity matrix of four cancer types}
	\label{fig:eigenValues_4cancers}
\end{figure}
 
The fused patient affinity matrix $\mathbf{W}$ generated by ANF is a state transition matrix, and is asymmetric in most cases. We chose the affinity matrices that achieve best clustering accuracies for four cancer types, and calculated the eigenvalues of the corresponding normalized graph Laplacian of these matrices (shown in Fig.~\ref{fig:eigenValues_4cancers}).
We found eigengap heuristic is very useful for deciding the number of clusters. For example, the first two smallest eigenvalues for adrenal gland are very near 0, while the third one is about 0.2. The eigengap between the second and third smallest values is relatively large. This suggests there should be two ``natural'' clusters (corresponding to the two nearly 0 eigenvalues). Furthermore, the eigengap between the fourth and third values is relatively high, too. This suggests we can use the learned affinity matrix for disease subtype discovery for adrenal gland cancer. In fact, when we set the number of clusters to be 3, our framework will separate 176 ``TCGA-PCPG'' samples into two groups consisting 155 samples and 21 samples respectively. If we set the number of clusters to 4, the 155 samples will be further split into two small groups as shown in Table.~\ref{tbl:confusion_adrenal}.

\begin{table}[t]
	\begin{center}
		\caption{Confusion matrix of clustering adrenal gland}
		\label{tbl:confusion_adrenal}
		\begin{tabular}{c|c|ccccc}
			\firsthline
			\#Clusters & TrueClass & \multicolumn{5}{c}{Clusters} \\
			\hline
			\multirow{3}{*}{2} & & C1 & C2 \\
			\cline{3-4}
			& TCGA-ACC & 0 & 76 \\ 
			\cline{3-4}
			& TCGA-PCPG & 176 & 1 \\
			\hline
			\multirow{3}{*}{3} & & C1 & C2 & C3\\
			\cline{3-5}
			& TCGA-ACC & 0 & 0 & 76\\ 
			\cline{3-5}
			& TCGA-PCPG & 155 & 21 & 1\\
			\hline
			\multirow{3}{*}{4} & & C1 & C2 & C3 & C4\\
			\cline{3-6}
			& TCGA-ACC & 0 & 0 & 76 & 0\\ 
			\cline{3-6}
			& TCGA-PCPG & 83 & 21 & 1 & 72\\
			\hline
			\multirow{3}{*}{5} & & C1 & C2 & C3 & C4 & C5\\
			\cline{3-7}
			& TCGA-ACC & 0 & 0 & 30 & 46 & 0\\ 
			\cline{3-7}
			& TCGA-PCPG & 83 & 21 & 0 & 1 & 72\\
			\lasthline
		\end{tabular}
	\end{center}
\end{table}

Analysis for other cancer types are similar and omitted here. With fused affinity matrix generated by ANF, users can calculate the eigenvalues of its normalized graph Laplacian, and use eigengap heuristic or more advanced techniques such as \cite{Wang2017} to determine the number of clusters.

\subsection{Neural Network Semi-supervised Learning}
In this section, we switch gear to examine the performance of a semi-supervised learning model that combines ANF and neural network. Fig.~\ref{fig:nn-classifier} shows the model architecture. We feed the output of ANF, i.e., kNN affinity matrices, to a neural network, and train the neural network with a few labeled examples (i.e., few-shot learning).

\subsubsection{Few-shot Learning: Training With A Few Examples}
We used the same architecture in Fig.~\ref{fig:nn-classifier} and  
parameter settings for all four cancer types. 
The numbers of hidden units in \textbf{Layer 0, 1, 2} are equal to the number of patients, 50, and the number of clusters (either 2 or 3), respectively. 
We used Adam optimizer with learning rate $= 0.05$, learning rate decay $= 0.9$ (decay learning rate every 10 iterations), maximum number of iterations $=100$. 

\paragraph{Training examples} 
Since we have true disease type information, we can randomly select a few examples with true class labels. We call these examples as \textbf{``clean''} examples.

Meanwhile, we use ANF Alg.~\ref{alg:ANF_framework} with uniform view weights to generate clustering results, and randomly select a few examples with cluster labels for training. We call these examples as \textbf{``noisy''} examples. 

In the following we randomly selected an incremental number of clean or noisy examples from less than 1\% to 50\% for training, and ran experiments five times and reported the best NMI (normalized mutual information) values for test sets (The training accuracy is 100\% for most of the times).

\begin{figure*}[t!]
	\centering
	\begin{subfigure}[t]{0.5\textwidth}
		\centering
		\includegraphics[height=1.65in]{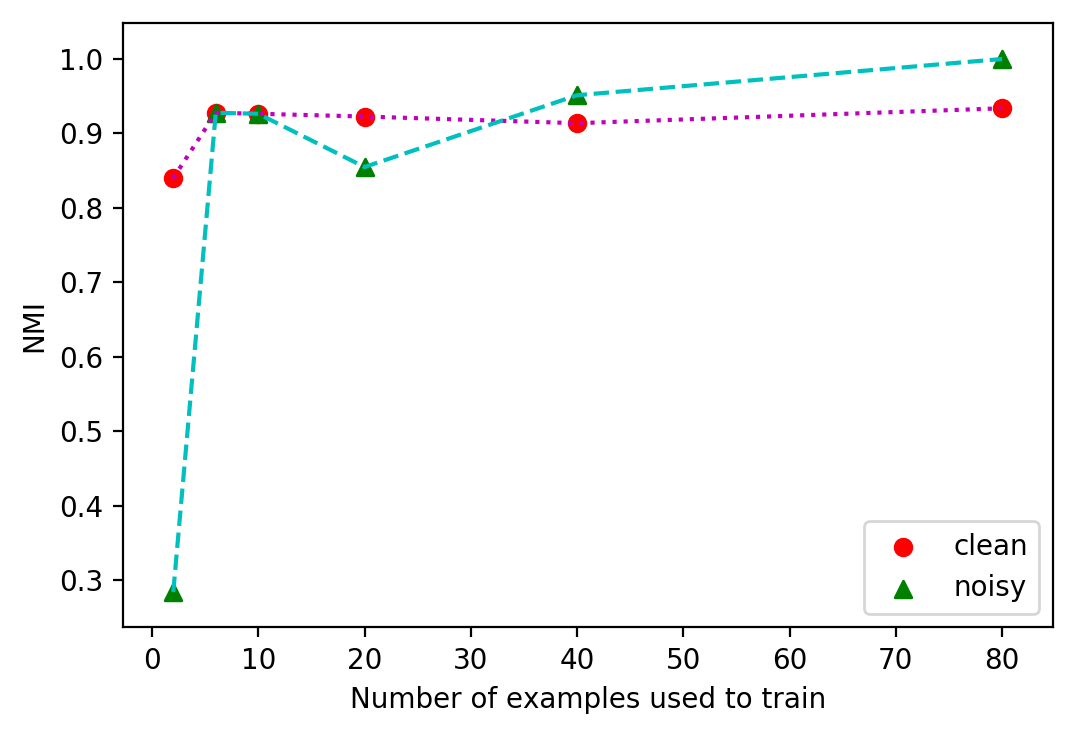}
		\caption{253 adrenal gland cancer samples}
	\end{subfigure}%
	~ 
	\begin{subfigure}[t]{0.5\textwidth}
		\centering
		\includegraphics[height=1.65in]{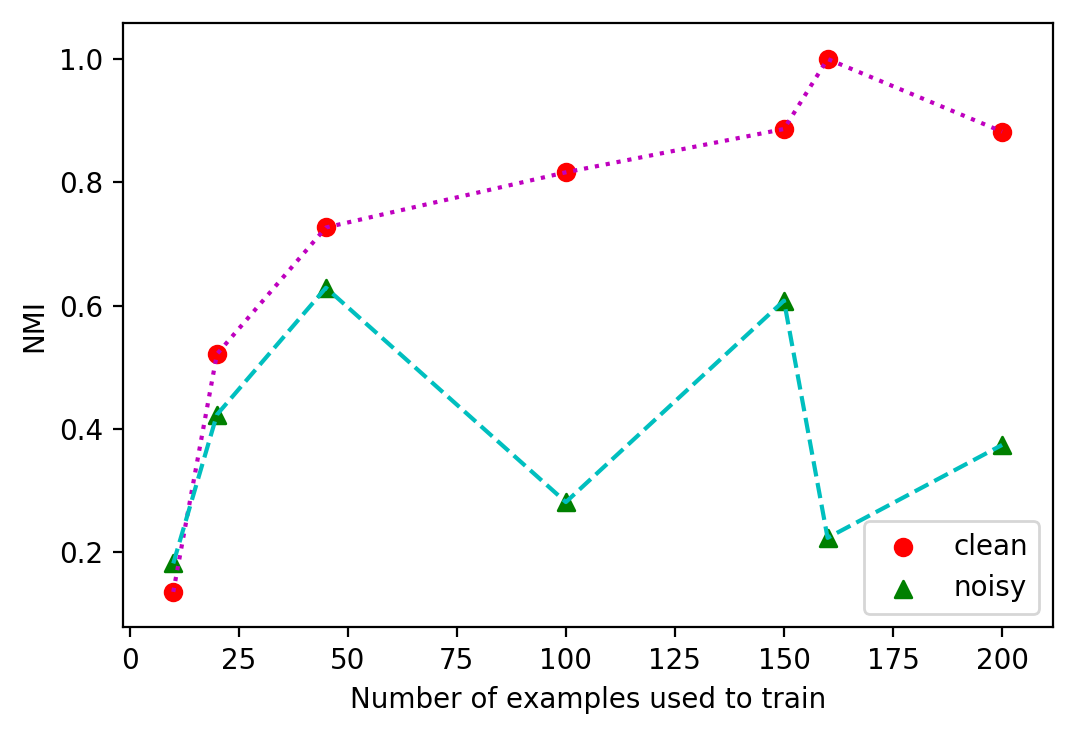}
		\caption{475 uterus cancer samples}
	\end{subfigure}
    ~
	\begin{subfigure}[t]{0.5\textwidth}
		\centering
		\includegraphics[height=1.65in]{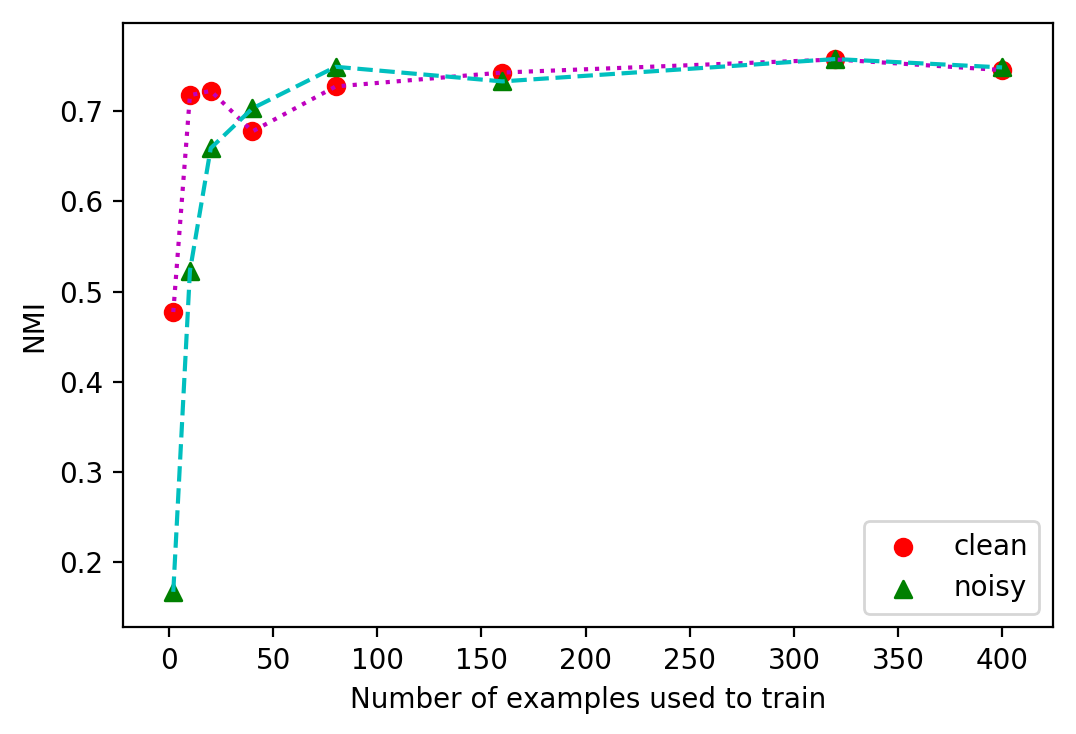}
		\caption{811 lung cancer samples}
	\end{subfigure}%
	~ 
	\begin{subfigure}[t]{0.5\textwidth}
		\centering
		\includegraphics[height=1.65in]{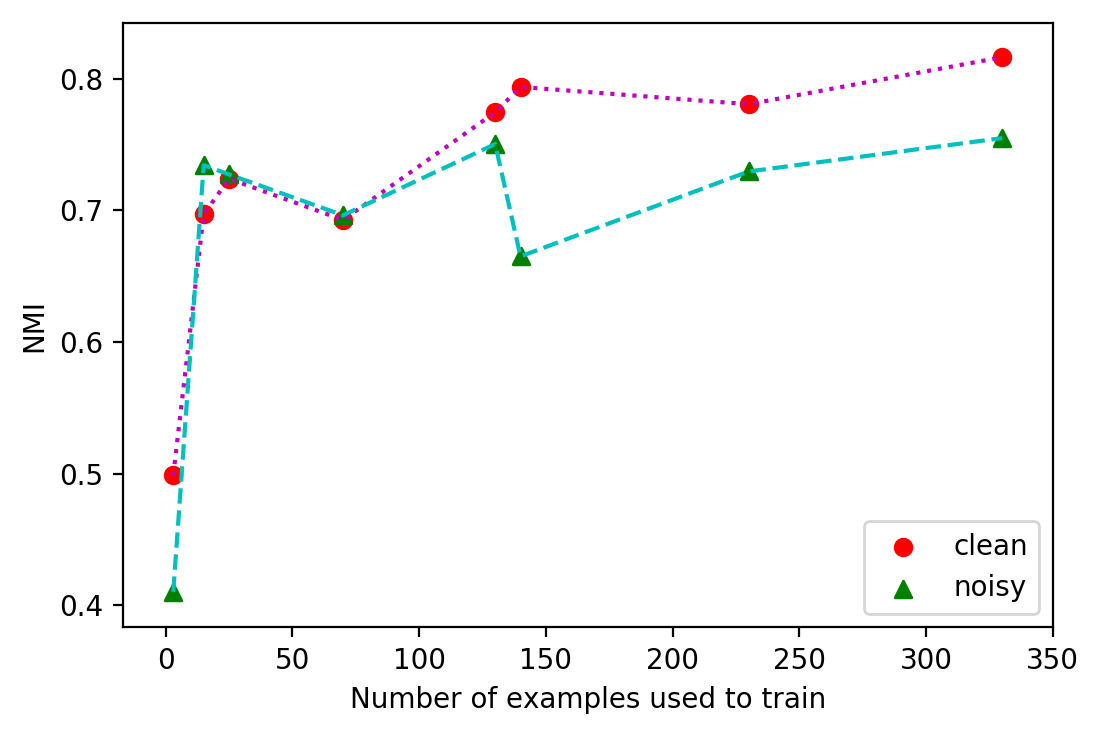}
		\caption{654 kidney cancer samples}
	\end{subfigure}
	\caption{Training NN classifier with very few examples}
	\label{fig:clean_vs_noisy}
\end{figure*}

Fig.~\ref{fig:clean_vs_noisy} shows that training with ``clean'' examples performs slightly better than ``noisy'' examples, especially for uterus and kidney cancer. Even ``noisy'' examples work well because that they are generated by unsupervised ANF and are in fact correct in most cases. Amazingly, training with only two clean examples for adrenal gland cancer achieves NMI $\ge$ 0.8 (accuracy $\ge$ 97\%). With only 10 training examples, the NMI value becomes larger than 0.9.  Overall as the number of training examples increases, the performance improves except training uterus cancer with noisy examples. 

Usually with few training examples and high dimensional features, it is not possible to achieve high test set accuracy. However, due to the good representation learned by Alg.~\ref{alg:ANF_framework} as the initial input for neural network, we can avoid overfitting and achieve surprisingly good results on test set with less than 1\% of training examples. 

\textbf{How Does the Classifier Work Internally?}

To see why training with only one example each class can achieve such high accuracy in adrenal gland cancer (results for other cancer types are similar and not shown here), we visualize the internal states of \textbf{Layer 1} and \textbf{Layer 2} both before and after training. \textbf{Layer 1} has 50 dimensions, we used PCA for dimensionality reduction. 

\begin{figure*}[t!]
	\centering
	\begin{subfigure}[t]{0.5\textwidth}
		\centering
		\includegraphics[height=2.2in]{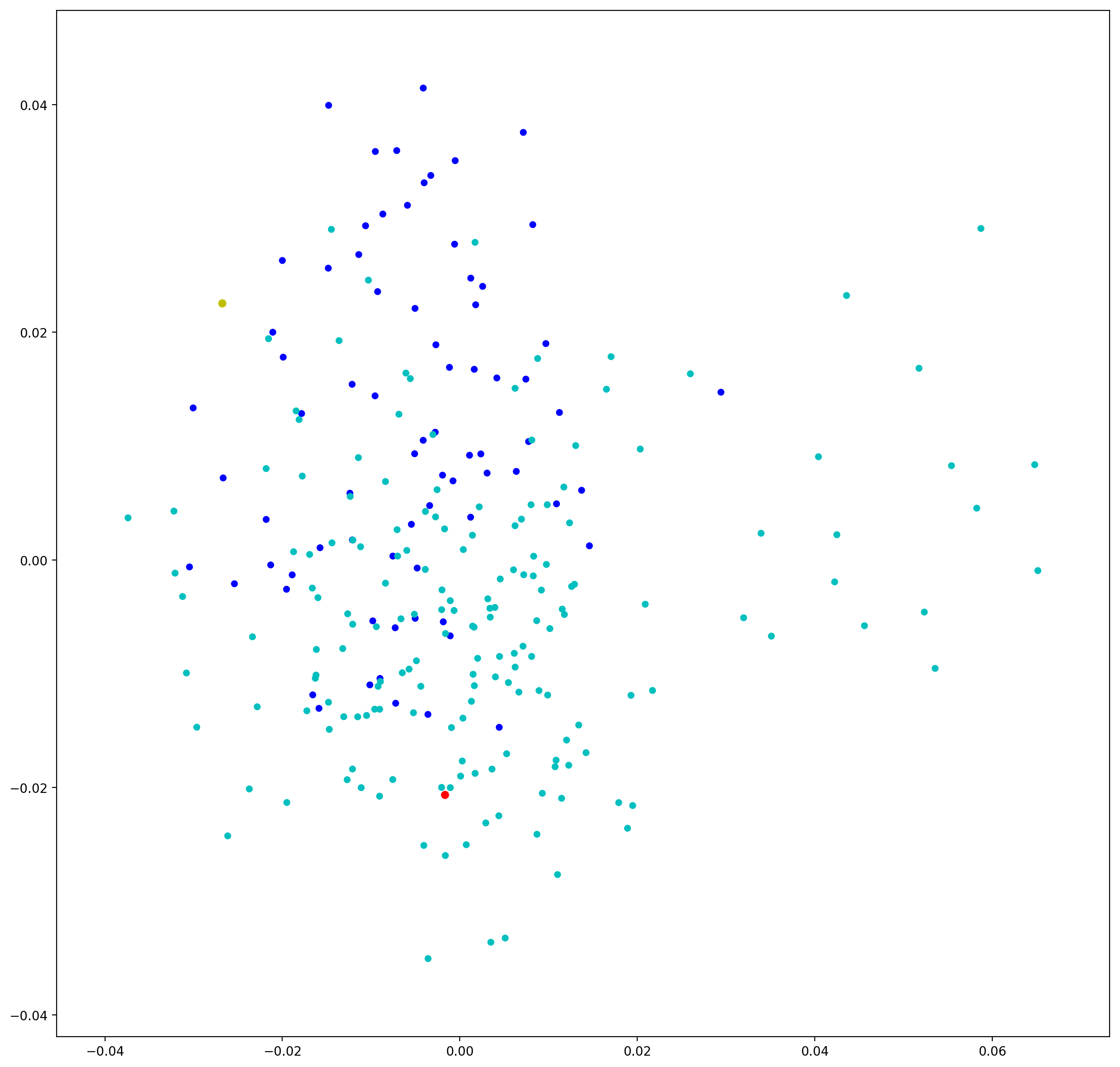}
		\caption{Before training: 2nd to last layer}
	\end{subfigure}%
	~ 
	\begin{subfigure}[t]{0.5\textwidth}
		\centering
		\includegraphics[height=2.2in]{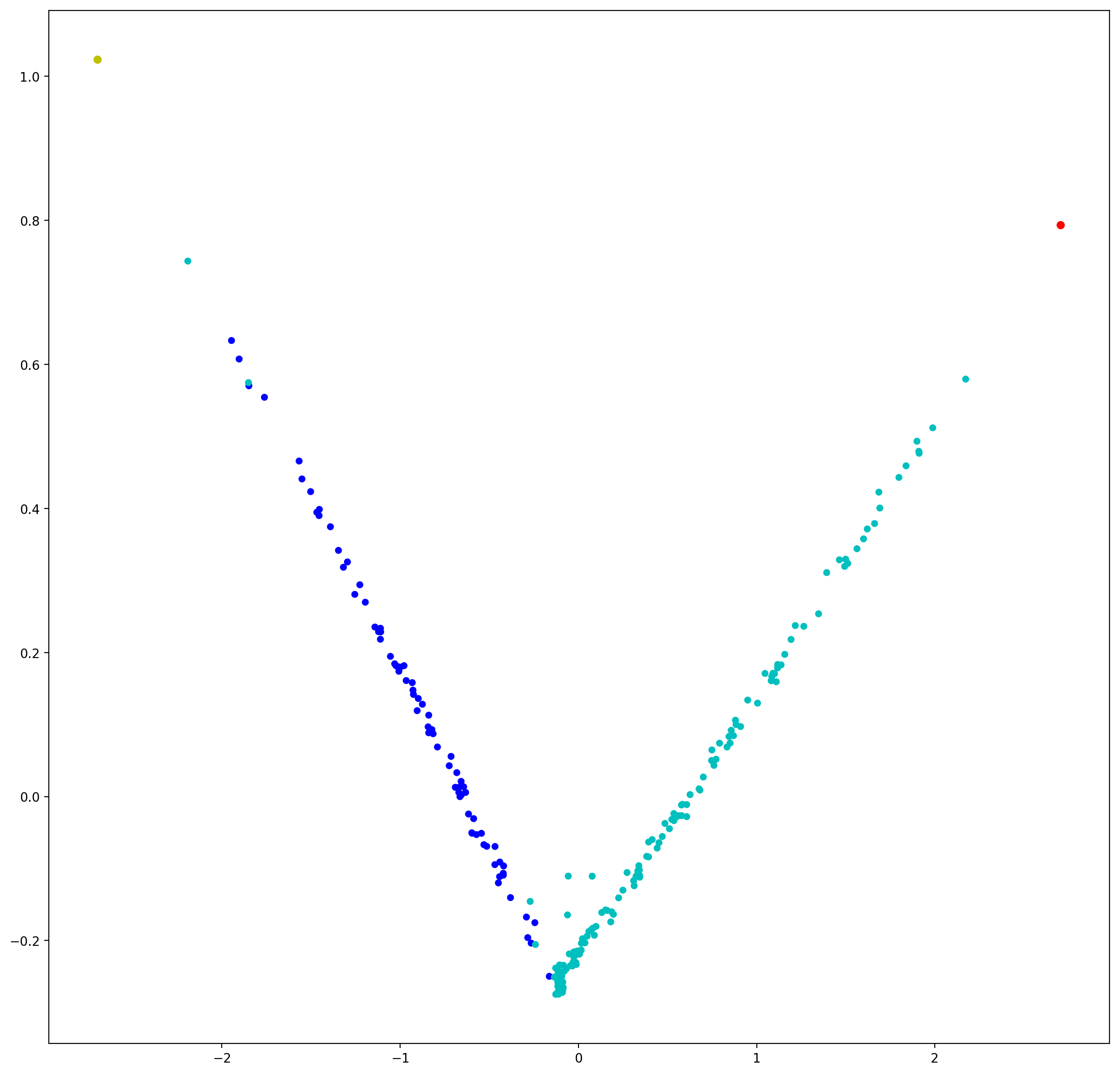}
		\caption{After training: 2nd to last layer}
	\end{subfigure}
	~
	\begin{subfigure}[t]{0.5\textwidth}
		\centering
		\includegraphics[height=2.2in]{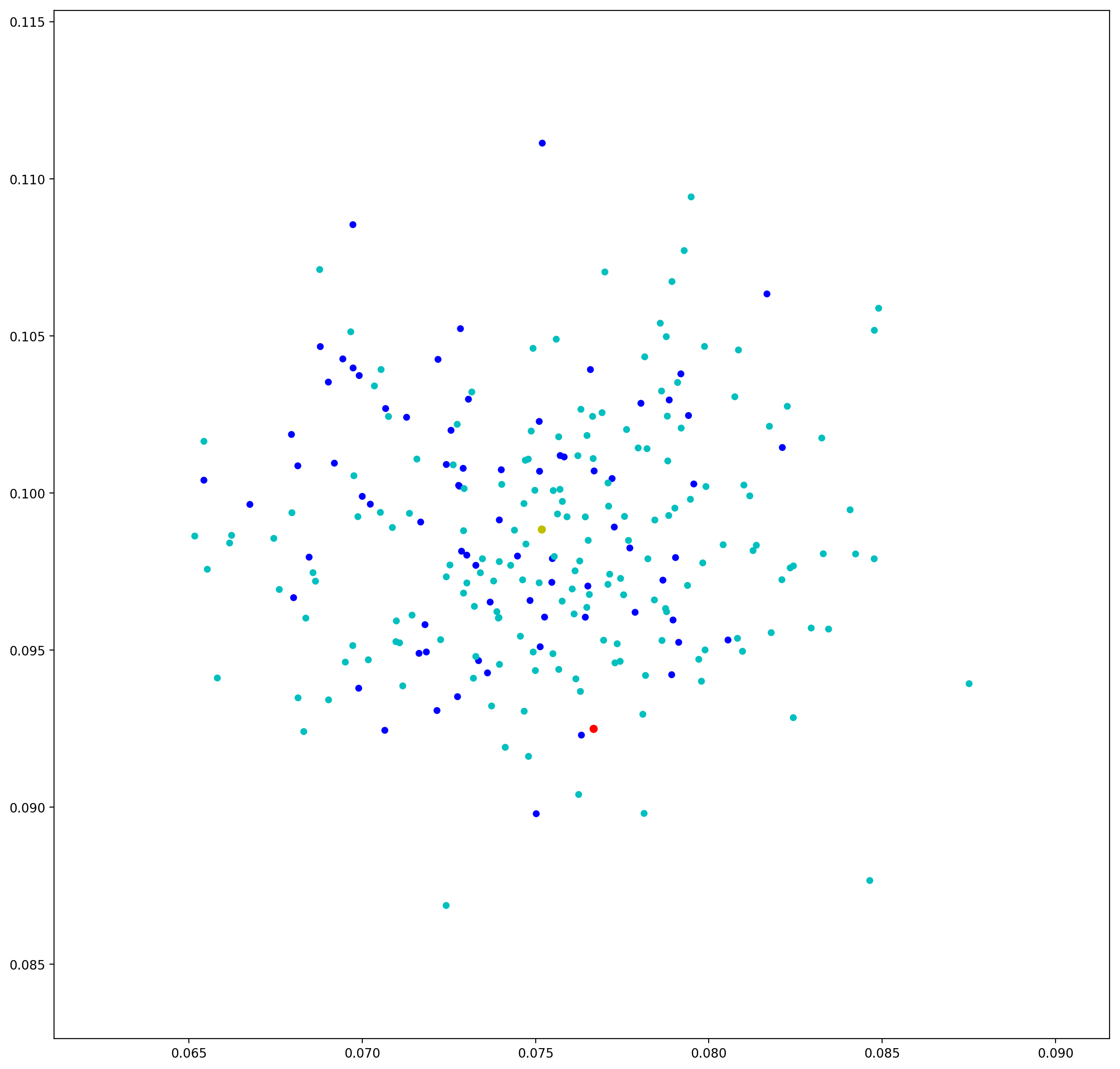}
		\caption{Before training: last layer}
	\end{subfigure}%
	~ 
	\begin{subfigure}[t]{0.5\textwidth}
		\centering
		\includegraphics[height=2.2in]{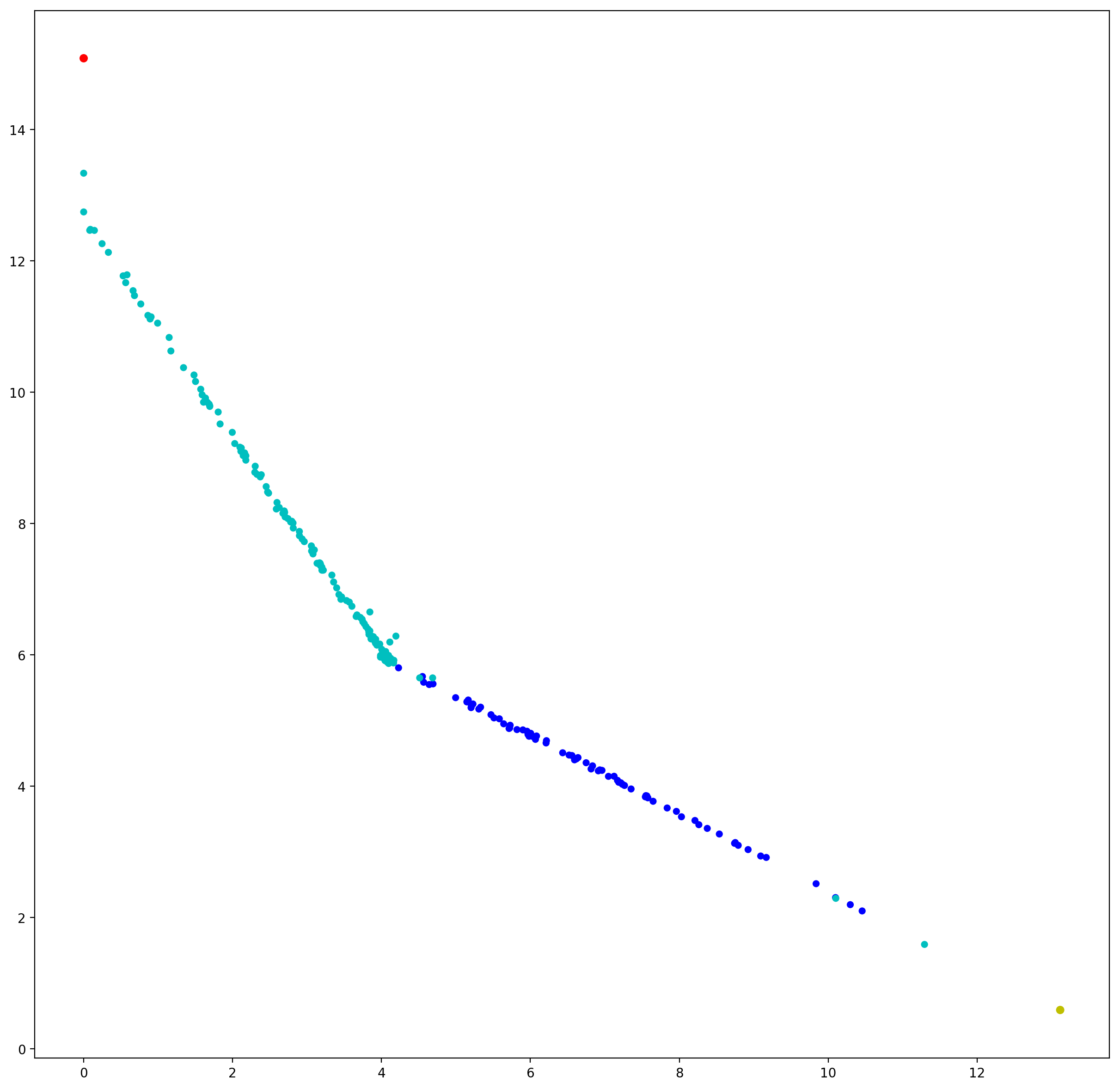}
		\caption{After training: last layer}
	\end{subfigure}
\caption{Internal states visualization. Training with one example per class for adrenal gland cancer achieves 97\% accuracy on test set}
	\label{fig:one_example_training}
\end{figure*}

In Fig.~\ref{fig:one_example_training}, each data point corresponds to a patient and is colored blue or cran corresponding to the true class label. The two examples used for training are colored yellow and red. As we can see, the state of second to last layer and the last layer are both random before training due to random initialization of model parameters. The two training examples ``drag'' all the samples of its class to its side during training. After training, we can clearly see two linearly separable clusters.

\subsubsection{Finetune Clustering Result}
Our framework not only facilitates few-shot learning, but can learn a good representation for transfer learning as well. We demonstrate this with the following experiments on kidney cancer dataset.

Kidney cancer has \textbf{three} disease types. We first select a few examples from only \textbf{two} disease types and train the model (Fig.~\ref{fig:nn-classifier}). Even though trained on only \textbf{two} disease types, we expect the model to be able to learn a good representation that may reflect three natural clusters. In order to test this, we ``freeze'' the model parameters in lower layers, and finetune the last layers of the model with both clean and noisy examples (which are again obtained through unsupervised ANF and spectral clustering) from all three disease types.
 
\begin{figure*}[t!]
	\centering
	\begin{subfigure}[t]{0.5\textwidth}
		\centering
		\includegraphics[height=1.65in]{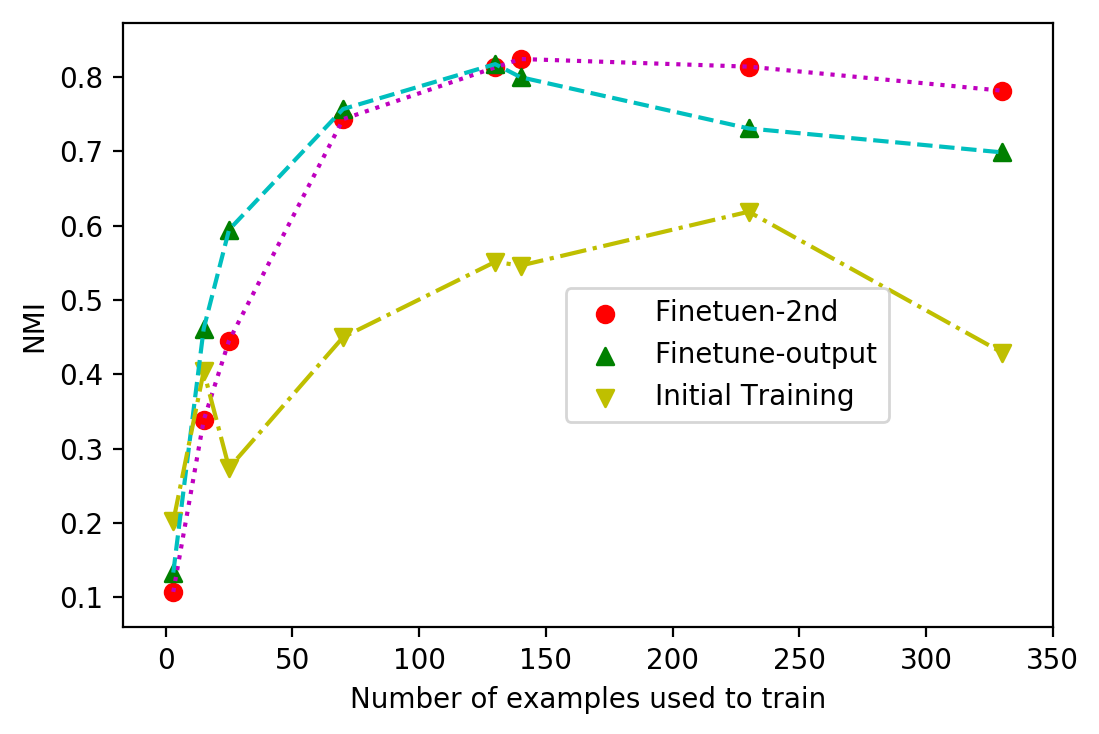}
		\caption{Train two views with clean examples}
	\end{subfigure}%
	~ 
	\begin{subfigure}[t]{0.5\textwidth}
		\centering
		\includegraphics[height=1.65in]{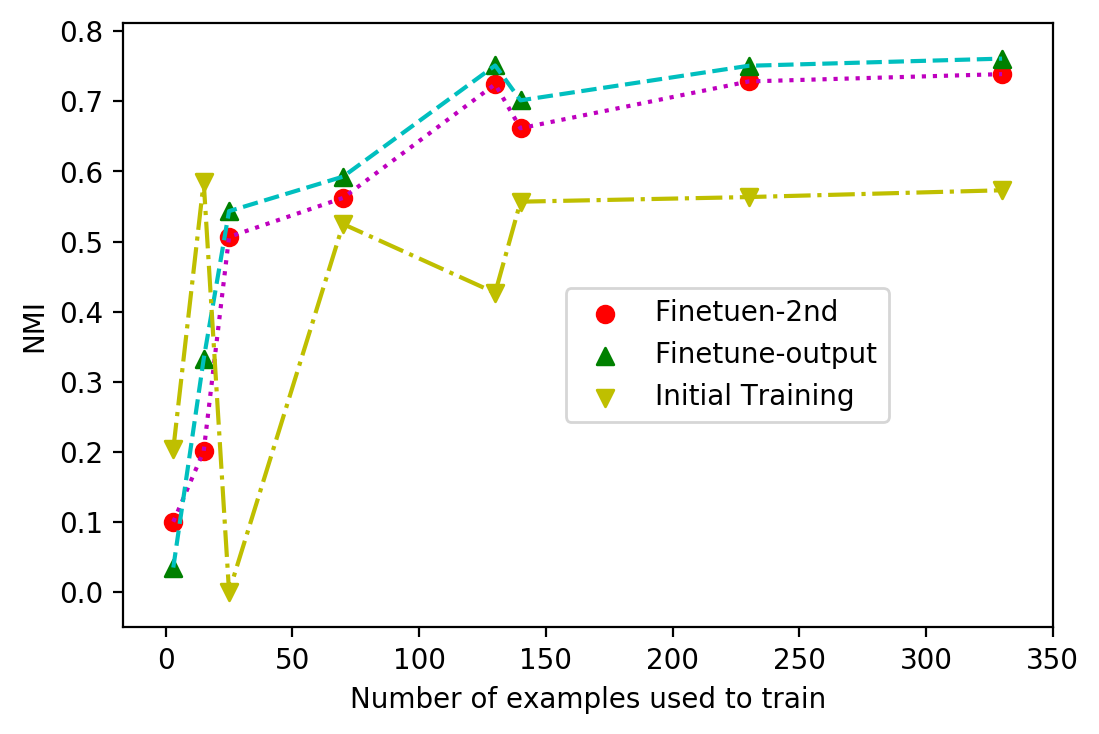}
		\caption{Train two views with noisy examples}
	\end{subfigure}
	~
	\begin{subfigure}[t]{0.5\textwidth}
		\centering
		\includegraphics[height=1.65in]{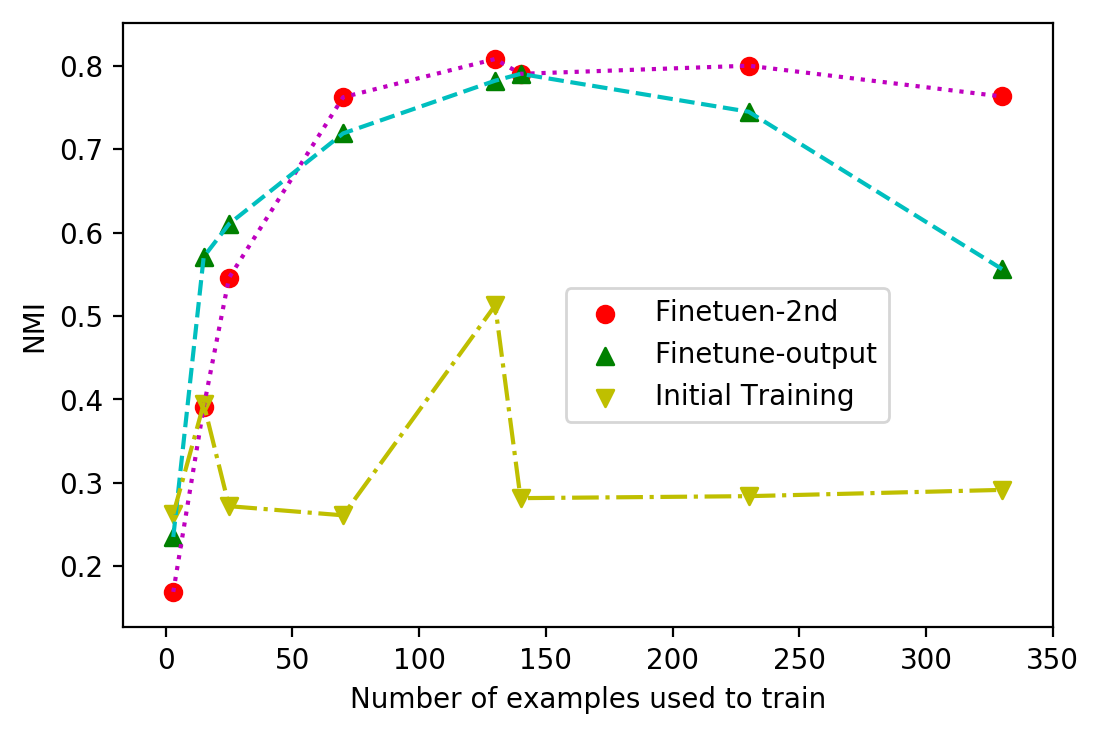}
		\caption{Train three views with clean samples}
	\end{subfigure}%
	~ 
	\begin{subfigure}[t]{0.5\textwidth}
		\centering
		\includegraphics[height=1.65in]{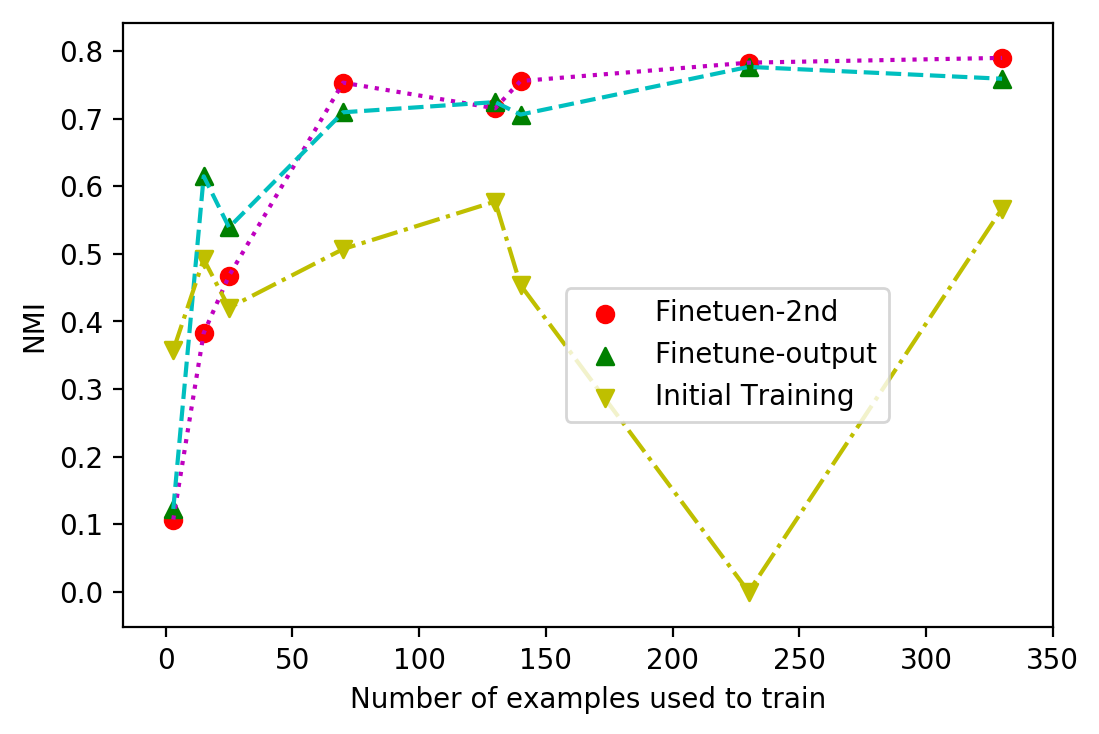}
		\caption{Train three views with noisy examples}
	\end{subfigure}
	\caption{Finetune Classifier for Kidney Cancer}
	\label{fig:finetune_kidney}
\end{figure*}

Fig.~\ref{fig:finetune_kidney} shows that finetuning model significantly improves performance after initially training with examples from only two disease types (downward yellow triangle in the figure). This means the learned patient representation (which has been ``freezed'' during finetuning) is good enough to reflect the three true clusters even though we only trained on two clusters for learning a representation. 

In addition, we trained our model using two views (``gene+mirnas'', results shown in the upper panel of Fig.~\ref{fig:finetune_kidney}) and three views (``gene+mirnas+methylation'', results shown in the lower panel of Fig.~\ref{fig:finetune_kidney}), and finetuned the last layer only (upward green triangle in Fig.~\ref{fig:finetune_kidney}) and the last two layers (filled red circle in Fig.~\ref{fig:finetune_kidney}). We found that finetuning the second to last layer performs slightly better than finetuning only the last output layer, especially when there are more views (as shown in the lower left panel of Fig.~\ref{fig:finetune_kidney}), suggesting the second to last layer may be more useful for transfer learning. 

\section{Discussion and Conclusion}
Defining cancer subtypes and identifying subtype-specific molecular signatures associated with clinical variables is one major goal for cancer genomics. 
In this paper, we presented both unsupervised and semi-supervised affinity network fusion (ANF) framework that can integrate multi-omic data for cancer patients clustering and subtype discovery. 

We used the newest release of harmonized cancer datasets from Genomic Data Commons Data Portal (since the data is harmonized, it is of high quality for large-scale integration), and selected 2193 cancer patients from four primary sites with known disease types. 
The experimental results on this relatively large dataset (2193 cancer patients with gene expression, miRNA expression and DNA methylation data) are very promising. The learned fused affinity matrices for the selected four cancer types matched well with both true class labels and the theory of spectral clustering based on eigengap analysis, which can be reliably used for unknown cancer subtype discovery and identifying subtype-specific molecular signatures. 

While ANF itself can be used for unsupervised, zero-shot learning (i.e., cluster patients without any training examples), we further developed a semi-supervised learning model combining ANF and neural network, which achieved very good results for few-shot learning (e.g., being able to achieve 97\% accuracy on test set with training less than 1\% of data for classifying patients into correct disease types). This results from the good representation learned by ANF through effective kNN-based nonlinear transformations that reduce noise in multi-omic data. Our work shows the potential for combining supervised and unsupervised learning through good representation learning. 

While we only reported experimental results on four cancer types with known disease types, ANF can be used for discovering subtypes of other cancers, and more generally, for complex object clustering with multi-view feature matrices.

\section*{Funding}
This work was supported in part by the US National Science Foundation under grants NSF IIS-1218393 and IIS-1514204. Any opinions, findings, and conclusions or recommendations expressed in this material are those of the author(s) and do not necessarily reflect the views of the National Science Foundation.



\end{document}